\documentclass[10pt]{article}
\usepackage{amsmath}
\usepackage{amssymb}
\usepackage{graphicx}
\usepackage{cite}
\usepackage{color} 
\usepackage[labelfont=bf,labelsep=period,justification=raggedright]{caption}
\makeatletter
\renewcommand{\@biblabel}[1]{\quad#1.}
\makeatother

\date{}

\begin{document}

\begin{flushleft}
{\Large
\textbf{Evolution of interactions and cooperation in the spatial prisoner's dilemma game}
}\sffamily
\\[3mm]
\textbf{Chunyan Zhang,$^{1,2,\ast}$ Jianlei Zhang,$^{1,2}$ Guangming Xie,$^{1,\dagger}$ Long Wang,$^{1}$ Matja{\v z} Perc$^{3,\S}$}
\\[2mm]
{\bf 1} State Key Laboratory for Turbulence and Complex Systems, College of Engineering, Peking University, Beijing, China
{\bf 2} Theoretical Biology Group, Centre for Ecological and Evolutionary Studies, University of Groningen, Kerklaan, The Netherlands
{\bf 3} Department of Physics, Faculty of Natural Sciences and Mathematics, University of Maribor, Slovenia
\\[1mm]
$^{\ast}$zhcy@pku.edu.cn\\
$^{\dagger}$xiegming@pku.edu.cn\\
$^{\S}$matjaz.perc@uni-mb.si [www.matjazperc.com]
\end{flushleft}
\sffamily
\section*{Abstract}
We study the evolution of cooperation in the spatial prisoner's dilemma game where players are allowed to establish new interactions with others. By employing a simple coevolutionary rule entailing only two crucial parameters, we find that different selection criteria for the new interaction partners as well as their number vitally affect the outcome of the game. The resolution of the social dilemma is most probable if the selection favors more successful players and if their maximally attainable number is restricted. While the preferential selection of the best players promotes cooperation irrespective of game parametrization, the optimal number of new interactions depends somewhat on the temptation to defect. Our findings reveal that the ``making of new friends'' may be an important activity for the successful evolution of cooperation, but also that partners must be selected carefully and their number limited.

\section*{Introduction}
Social dilemmas are situations in which the optimal decision for an individual is not optimal, or is even harmful, for the society as a whole. Rational agents, who seek to maximize their own wellbeing, may thus attempt to free ride and reap undeserved rewards, \textit{i.e.} benefit from the ``social'' contributions of others without providing their own in exchange. However, many simple as well as complex organisms, including higher mammals and humans, exhibit a large tendency towards altruistic behavior. Resolving a social dilemma entails providing a rationale on how can behavior that is costly for an individual but beneficial for the society be maintained by means of natural selection? Achieving a satisfactory understanding of the evolution of cooperation in situations constituting a social dilemma is in fact fundamental for elucidating and properly comprehending several key issues that humanity is faced with today, including sustainable management of environmental resources and warranting satisfactory social benefits for all involved, to name but a few.

Evolutionary game theory has a long and very fruitful history when it comes to understanding the emergence and sustainability of cooperative behavior amongst selfish and unrelated individuals at different levels of organization. Several comprehensive books \cite{axelrod_84, sigmund_93, weibull_95, hofbauer_98, gintis_00, nowak_06, sigmund_10} and reviews \cite{doebeli_el05, nowak_s06, szabo_pr07, schuster_jbp08, perc_bs10} are available that document the basics as well as past advances in a cohesive and readily accessible manner. The prisoner's dilemma game in particular is frequently employed for studying the evolution of cooperative behavior among selfish individuals. In it's original form, the prisoner's dilemma game consists of two players who have to decide simultaneously whether they wish to cooperate or to defect. The dilemma is given by the fact that although mutual cooperation yields the highest collective payoff, which is equally shared among the two players, individual defectors will do better if the opponent decides to cooperate. Since selfish players are aware of this fact they both decide to defect, whereby none of them gets a profit. Thus, instead of equally sharing the rewarding collective payoff received by mutual cooperation, they end up empty-handed.

A key observation in recent history related to the resolution of the prisoner's dilemma game was that spatial reciprocity can maintain cooperative behavior without any additional assumptions or strategic complexity \cite{nowak_n92b} (see also \cite{hauert_n04}). Other well known mechanisms promoting cooperation include kin selection \cite{hamilton_wd_jtb64a}, direct and indirect reciprocity \cite{axelrod_s81, ohtsuki_jtb04, ohtsuki_jtb06b, brandt_jtb06, pacheco_jtb08}, as well as group \cite{wilson_ds_an77, dugatkin_bs96} and multilevel selection \cite{traulsen_pnas06, szolnoki_njp09}. These as well as related mechanism for the promotion of cooperation have been comprehensively reviewed in \cite{nowak_s06}. Another important development that facilitated the understanding of the evolution of cooperation came in the form of replacing the initially employed regular interaction graphs, \textit{e.g.} the square lattice, with more complex networks \cite{abramson_pre01, ebel_pre02, holme_pre03, wu_zx_pre05, tomassini_pre06, ohtsuki_n06, ren_pre07, wu_zx_pa07, vukov_pre08, luthi_pa08, floria_pre09, poncela_epl09}, whereby in particular the scale-free network has been identified as an excellent host topology for cooperative individuals \cite{santos_prl05, santos_pnas06}, warranting the best protection against the defectors. Since the strong heterogeneity of the degree distribution of scale-free networks was identified as a key driving force behind flourishing cooperative states \cite{santos_jeb06, gomez-gardenes_prl07, poncela_njp07, szolnoki_pa08, gomez-gardenes_jtb08}, some alternative sources of heterogeneity were also investigated as potential promoters of cooperation with noticeable success. Examples of such approaches include the introduction of preferential selection \cite{wu_zx_pre06}, asymmetry of connections \cite{kim_bj_pre02}, different teaching capabilities \cite{szolnoki_epl07}, heterogeneous influences \cite{wu_zx_cpl06}, social diversity \cite{perc_pre08} as well as diversity of reproduction time scales \cite{wu_zx_pre09b}. Evolutionary games on graphs have recently been comprehensively reviewed in \cite{szabo_pr07}, while related coevolutionary games have been reviewed in \cite{perc_bs10}. Comprehensive reviews concerning complex networks, on the other hand, include \cite{albert_rmp02, newman_siamr03, dorogovtsev_03, boccaletti_pr06}.

Coevolutionary games in particular have also received substantial attention recently, for example when studying the coevolution of strategy and structure \cite{pacheco_prl06}, games on networks subject to random or intentional rewiring procedures \cite{ebel_pre02, zimmermann_pre04, zimmermann_pre05, perc_njp06c, szolnoki_epl09, fu_pre09, wu_b_pone10}, prompt reactions to adverse ties \cite{van-segbroeck_bmceb08, van-segbroeck_prl09}, games on growing networks \cite{poncela_ploso08, poncela_njp09}, multiadaptive game \cite{lee_s_prl11}, and indeed many more \cite{vainstein_jtb07, szolnoki_njp08, szolnoki_epjb09, droz_epjb09, wardil_epl09, wardil_pre10, szabo_epl09, szolnoki_pre09, szolnoki_pre09d, sicardi_jtb09, tanimoto_pa09, cao_l_jtb11}. Here we aim to elaborate on this subject further by studying the evolution of cooperation in the prisoner's dilemma game where players are allowed to form new connections with other players that are not in their immediate neighborhoods. Conceptually the study is similar to \cite{szolnoki_epl08}, where it has been reported that the making of new connections promotes cooperation and may help resolve social dilemmas, yet here we focus more precisely on the impact of preference towards linking together more successful players (as opposed to just randomly selecting individuals to connect), as well as on the impact of the number of new links. For this we adopt the linking procedure proposed in \cite{poncela_ploso08}, but do not allow new players to join, \textit{i.e.} the network does not grow in size. Initially every player is connected only to its four nearest neighbors, and subsequently, at fixed time intervals, $m$ new links are introduced amongst players. Whether more successful players are more likely to receive a new link is determined by a single parameter $\lambda \in [0,1]$, whereby $\lambda \rightarrow 0$ gives all players equal chances (the introduction of new links is independent of the evolutionary success of individuals), while $\lambda \rightarrow 1$ strongly favors the more successful. All the details of the considered setup are described in the Methods section, while here we proceed with presenting the main results.

\section*{Results}
We start revealing the properties of the introduced model by examining the impact of the number of newly added links $m$ at each full iteration on the fraction of cooperators within the employed prisoner's dilemma game. Figure~1 shows the results obtained by a given combination of the temptation to defect $b$ and the parameter $m$. Apparently, the density of cooperators depends strongly on $m$. While the fraction of cooperators decreases monotonously from $1$ (\textit{i.e.} a state of full cooperation) to $0$ as $b$ increases, this transition occurs at different values of $b$ depending on $m$. It can be observed that the cooperative behavior is promoted for small and intermediate values of $m$, but as the parameter $m$ is increased further and exceeds a threshold value (approximately $m=3$), the system undergoes a transition in which the cooperation-facilitative effect deteriorates. These results indicate that an optimal value of $m$ warranting the most significant benefits to cooperators exists. Results presented in Fig.~1 evidence that there exist an optimal amount of new interactions to be added at each full iteration step, determined by $m$ via the coevolutionary process, for which the density of cooperators is enhanced best. It can be argued that for low values of $m$ (\textit{e.g.} $m=1$ in Fig.~1) the number of newly added links at each iteration is too small to allow the formation of strong hubs, which however, can emerge (see below) if the value of $m$ is sufficiently large (\textit{e.g.} $m=3$ in Fig.~1), yet not too large (\textit{e.g.} $m=9$ in Fig.~1). It is reasonable to expect that in the optimal case the degree distribution exhibits a heterogeneous outlay (see further below), in particular since such interaction networks are known to promote the evolution of cooperation \cite{santos_prl05}. Thus, high levels of cooperation are possible even at large $b$, as presented in Fig.~1. However, with $m$ exceeding the optimal value, the chosen players will establish many more connections, too many in fact, thereby essentially reducing the heterogeneity of the resulting interaction network and leaving the whole population in a state characterized by high connectivity resembling well-mixed conditions. Note that in well-mixed populations cooperators cannot survive if $b>1$, which explains why at large values of $m$ the evolution of cooperation in our case is less successful than at intermediate values of $m$.

The parameter $\lambda$ may also significantly affect the outcome of the game. In particular, larger values of $\lambda$ make it more likely for successful players (the ones with high payoffs) to become the recipients of new links. Results in Fig.~2 depict the average level of cooperation $f_{c}$ in dependence on the whole relevant span of the temptation to defect $b$ for different values of $\lambda$. It can be observed that at a fixed value of $b$ the presently studied model is increasingly more successful by promoting the evolution of cooperation as $\lambda$ increases. This is somewhat surprising as defectors will be the more successful players at least in the early stages of the game (when there are still enough cooperators to exploit), and thus one could further expect that by obtaining additional links they could outperform cooperators completely. Yet this is not what happens, and indeed when the probability to attach new links to the successful players is large (\textit{e.g.} $\lambda=0.99$ in Fig.~2), the cooperators can remain strong in numbers even if the temptation to defect is high. Based also on previous results \cite{szolnoki_epl08}, it is reasonable to conclude that high values of $\lambda$ promote the occurrence of a negative feedback effect that is associated with the defective but not with the cooperative behavior. Despite of the fact that initially (in early stages of the game) defectors can successfully extend their base of partners, ultimately their exploitative nature will convert all of them to defectors, and hence there will be nobody left to exploit. Such defector hubs are then quite vulnerable (in terms of the game they are unsuccessful), and are easily overtaken by cooperators. Once cooperators occupy such hubs, their mutually rewarding behavior strengthens their positions quickly, which ultimately paves the way for a successful evolution of cooperation that is here additionally promoted by the coevolutionary process of ``making new friends''.

Since networks are to be seen as evolving entities that may substantially affect the game dynamics that is taking place on them, it is also important to inspect the degree distribution of players in the employed system for different values of the temptation to defect $b$ as well as the clustering coefficient associated with the evolved networks. From the results presented in Fig.~3 it follows that the clustering coefficient of the initial square lattice (which is $0$) increases due to the addition of new links. This indicates that some realizations (depending on $\lambda$ and $b$) of the coevolutionary game give rise to compact clusters of players. By focusing first on the impact of $\lambda$, it can be observed that larger values promote clustering, albeit this depends also on the temptation to defect $b$. Especially in strongly defection-prone environments the larger values of $\lambda$ increase the clustering coefficient significantly. Since the parameter $\lambda$ controls the weight (\textit{i.e.} importance) of the payoffs during the coevolutionary process (the addition of new links), these results can be understood well. In particular, for small values of $\lambda$ the selection of players that will receive new links is virtually independent of the outcome of the game. In fact, all players are equiprobable recipients of new links, and hence the clustering coefficient is independent of $b$. On the other hand, larger values of $\lambda$ render the selection of the more successful players to become the recipients of new links more likely. From the degree distributions (not shown), we found that larger values of $\lambda$ lead to substantially more heterogeneous networks than small $\lambda$. Accordingly, the highly connected nodes are those successful players who accumulate higher payoffs, in turn receiving more and more new links if $\lambda \to 1$. This scenario holds virtually irrespective of $b$, only that for strong temptations to defect the clusters of cooperative players become larger, and accordingly larger is also the clustering coefficient presented in Fig.~3. As is traditionally argued, players located in the interior of such clusters enjoy the benefits of mutual cooperation and are therefore able to survive despite the exploitation from defectors. At this point we can conclude that high values of $\lambda$ enable cooperative players to grow relatively compact (well clustered) communities starting from their initial nearest neighbors, which in turn strongly promotes the evolution of cooperation, as evidenced by the results presented thus far.

With the aim of further enhancing our understanding of the presented results, we investigate this model also from the microscopic point of view, first by showing the fraction of new links received by cooperators in Fig.~4, and second by comparing the average payoffs of cooperators and defectors in Fig.~5. From the results presented in Fig.~4 two regimes can roughly be distinguished. For small values of $b$ large values of $m$ are optimal for cooperators to become the recipients of new links. When going towards larger $b$, however, there is a crossover, where finally for large temptations to defect intermediate values of $m$ emerge clearly as optimal for cooperators to receive at least some of the ``coevolutionary'' added links. These observations resonate with the preceding results (see Fig.~1), where indeed intermediate values of $m$ were found to be optimal for the evolution of cooperation, especially at large values of $b$. A relative straightforward view into the microscopic workings of the coevolutionary process reveals that this may in fact be because cooperators, despite of their inherent disadvantage over defectors, are still able to acquire at least some fraction of the newly introduced links between players if the value of $m$ is neither too small nor too large.

Results presented in Fig.~5 lend additional support to those presented in Fig.~4, which is expected since indeed if $\lambda \to 1$ the awarding of new links depends primarily on the payoffs of players. It can be observed that for small values of $b$ large values of $m$ ensure that the average payoff of cooperators is the highest if compared to the average payoff of defectors. When approaching larger $b$, however, there is again a crossover clearly inferable, such that only intermediate values of $m$ warrant cooperators to outperform defectors in terms of the average payoff. It may come as a surprise that despite of the fact that at $b=1.6$ the minority of players is adopting the cooperative strategy (even under optimal conditions in terms of $m$ and $\lambda$) their average payoff is still larger than that of the dominating defectors. After inspecting the distribution of strategies on the network in search for an explanation, we find that even under such unfavorable conditions in small isolated regions of the network the cooperators are surrounded by other cooperators in a very compact manner. Note that the clustering coefficient in this parameter range is relatively large, hence supporting the local formation of such cooperative clusters, in turn warranting a relatively high average payoff for the small population of cooperators. Nevertheless, the cooperators are unable to spread but can only maintain their existence within these clusters that emerge as a sort of a refuge due to the coevolutionary addition of new links, thereby protecting the cooperators from otherwise inevitable extinction.

Lastly, we also address briefly the issue of the importance of the initial state on the evolution of cooperation in the presently studied model. In Fig.~\ref{fig6} we present the fraction of cooperators in dependence on $b$ for different values of $\rho_{c}$, whereby $0 \leq \rho_{c} \leq 1$ is the fraction of cooperators in the whole population at the beginning of the game. All the results were obtained for $\lambda=0.99$, where the addition of new links is driven primarily by the payoff values that the individual players are able to acquire. It is interesting to observe that the initial strategy configuration in the population plays quite an important role. First, it is worth emphasizing the positive aspect, which is that cooperative behavior can ultimately be maintained even when $\rho_{c}$ is small (\textit{e.g.} $\rho_{c}=0.2$ in Fig.~6). Expectedly, for larger values of $\rho_{c}$ (\textit{e.g.} $\rho_{c}=0.6$ in Fig.~6) the evolution of cooperation is more robust, resulting in complete cooperator dominance over a significantly wider range of the temptation to defect $b$. However, with $\rho_{c}$ increasing further (\textit{e.g.} $\rho_{c}=0.8$ in Fig.~6), the defectors will recapture some advantages, and it becomes obvious that larger values of $\rho_{c}$ decrease the potentially constructive effect of coevolution on the promotion of cooperation within the present setup. Hence, we arrive at the conclusion that in the long run there is a maximal fraction of cooperators attainable only at an intermediate value of $\rho_{c}$.

\section*{Summary}
We have studied the evolution of cooperation in the spatial prisoner's dilemma game where players are allowed to establish new interactions with other players that are not necessarily within their immediate neighborhoods. While the question of whether new links amongst players may potentially promote cooperation has been addressed before \cite{szolnoki_epl08, poncela_ploso08, poncela_njp09}, we have here reexamined this by focusing more precisely on the impact of preference towards linking together more successful players (as opposed to just randomly selecting individuals to connect), as well as on the impact of the number of new links. In order to achieve this, we have adopted the linking procedure proposed in \cite{poncela_ploso08}, but did not allow the network of players to grow in size. We have found that the resolution of the social dilemma, here modeled by the prisoner's dilemma game, is most probable if the selection favors the more successful players and if the maximally attainable number of new links added to the population is restricted. More precisely, we have found that the more the selection favors the more successful players, the stronger the promotion of cooperation. Conversely, for the added number of new links it proved optimal if the latter is limited, although this conclusion depends somewhat also on the temptation to defect $b$. While for low values of $b$ a larger number of new links may be better, for high values of $b$ an intermediate number of new links is preferred. We have also examined the dependence of these results on the initial fraction of cooperators in the population, and found rather surprisingly that initially too highly cooperative states are not optimal starting points for the successful evolution of cooperation. We have argued that this may be due to the fact that defectors thrive in populations where there are numerous cooperators to exploit, and ultimately this may become a disadvantage in the latter stages of the game, although this observation may require additional research in order to be better understood. Altogether, our results indicate that new links amongst players may promote cooperation, although it is important to take into account many factors for this conclusion to remain valid. Most importantly, links should be established preferentially amongst the more successful players and must not be too many. This leads us to the reiteration of the statement from the Abstract of this paper, being that the ``making of new friends'' may be an important activity for the successful evolution of cooperation, but at the same time, it has to be emphasized that friends must be selected carefully and their number kept within reasonable bounds. We hope that this study will motivate further research on coevolutionary games and promote our understanding of the evolution of cooperation.

\section*{Methods}
We consider the spatial prisoner's dilemma game where each player occupies a node on the square lattice of size $N$ and is connected to its four nearest neighbors. Initially each player is designated either as a cooperator or defector with equal probability unless stated otherwise, and players obtain their payoffs by means of pairwise interactions with all their partners. Following standard practice, the payoffs are $T=b$ for a defector playing with a cooperator, $R=1$ for mutual cooperation, and $S=P=0$ for a cooperator facing a defector and mutual defection, respectively. We thus have the payoff matrix
\[\begin{array}{*{20}{c}}
   {} & {\begin{array}{*{20}{ll}}
   C & \ D  \\
\end{array}}  \\
   {\begin{array}{*{20}{c}}
   C \!\!\!\!\!\! \\
   D  \!\!\!\!\!\!\\
\end{array}} & {\left( {\begin{array}{*{20}{c}}
   {1} &\ {0}  \\
   b & \ 0  \\
\end{array}} \right)}  \\
\end{array}\]
with the only free parameter being the temptation to defect $b$. This setup preserve the essential dilemma in that no matter what the opponent does, defection leads to a higher (or at least equal) payoff. Selfish and rational players would therefore always choose defection. But since the payoff for mutual defection is smaller than the payoff for mutual cooperation ($R>P$) the dilemma arises on what to choose if having in mind also the welfare of the society and not just personal interests. As usual, in one full iteration cycle each agent plays the game once with all its neighbors.

Following payoff accumulation, players attempt to adopt strategies from their neighbors with the aim of increasing their fitness (success) in future rounds of the game. Suppose that player $x$ with $k_{x}$ neighbors (initially this will be four, but may increase due to coevolution) accumulates its payoff $p_{x}$. To update its strategy, player $x$ selects one player $y$ amongst its $k_{x}$ neighbors with equal probability ($=1/k_{x}$). Following \cite{szabo_pre98}, we use the Fermi strategy adoption function given by
\begin{equation}
W(s_{y} \rightarrow s_{x})=\frac{1}{1+\exp[(p_{x}-p_{y})/K]},
\end{equation}
which constitutes the probability that player $x$ will adopt the strategy of player $y$, where $K$ determines the uncertainty by strategy adoptions or its inverse the intensity of selection. In this work we set $K=0.1$, which strongly prefers strategy adoptions from the more successful players, yet it is not impossible that a player performing worse will be adopted either. All the players update their strategies according to this rule in a synchronous manner.

Importantly, here we extend the above traditional setup by allowing players to increase their neighborhoods by linking with players that may be far from their nearest neighbors. Thus, parallel with the evolution of strategies, interactions between players evolve as well. In particular, after every full iteration, $m$ new links are added amongst players while keeping the network size fixed at $N$. For every new link two individuals are chosen at random from the whole population, with the probability $Q_{i}(n)$ of choosing agent $i$ in game round $n$ defined as (following \cite{poncela_ploso08})
\begin{equation}
Q_{i}(n)=\frac{1-\lambda+\lambda
f_{i}(n)}{\sum_{j=1}^{N}[1-\lambda+\lambda f_{j}(n)]},
\end{equation}
where $N$ is the system size and $f_{j}(n)$ is the accumulated payoff of agent $j$. The parameter $\lambda \in [0,1]$ controls the importance of the payoffs in the creation of new links amongst players. The case of $\lambda=0$ corresponds to neutrality, where each player has equal chances of obtaining a new link, irrespective of its evolutionary success. Conversely, positive values of $\lambda$ render the selection of the more successful players more likely, \textit{i.e.} players with $f_{j}(n) \neq 0$ are chosen preferentially, while $\lambda=1$ implies that the selection probability is linear with the magnitude of the payoffs (indicating clearly that the most successful players are most likely to obtain new links). We emphasize that self-interactions and duplicate links are omitted. It is also important to note that the continuing addition of new links without growth, \textit{i.e.} new players, evidently leads to a fully connected network. Yet the time scales \cite{roca_prl06} in this model concerning the evolution of cooperation and the evolution of interactions are very different, such that a quasi stationary state of the two strategies is reached well before full connectedness. Since the focus here is on the evolution of cooperation, we stop the simulations once this quasi stationary state is reached to record the final results.

\clearpage

\begin{figure}
\begin{center}\includegraphics[width=9cm]{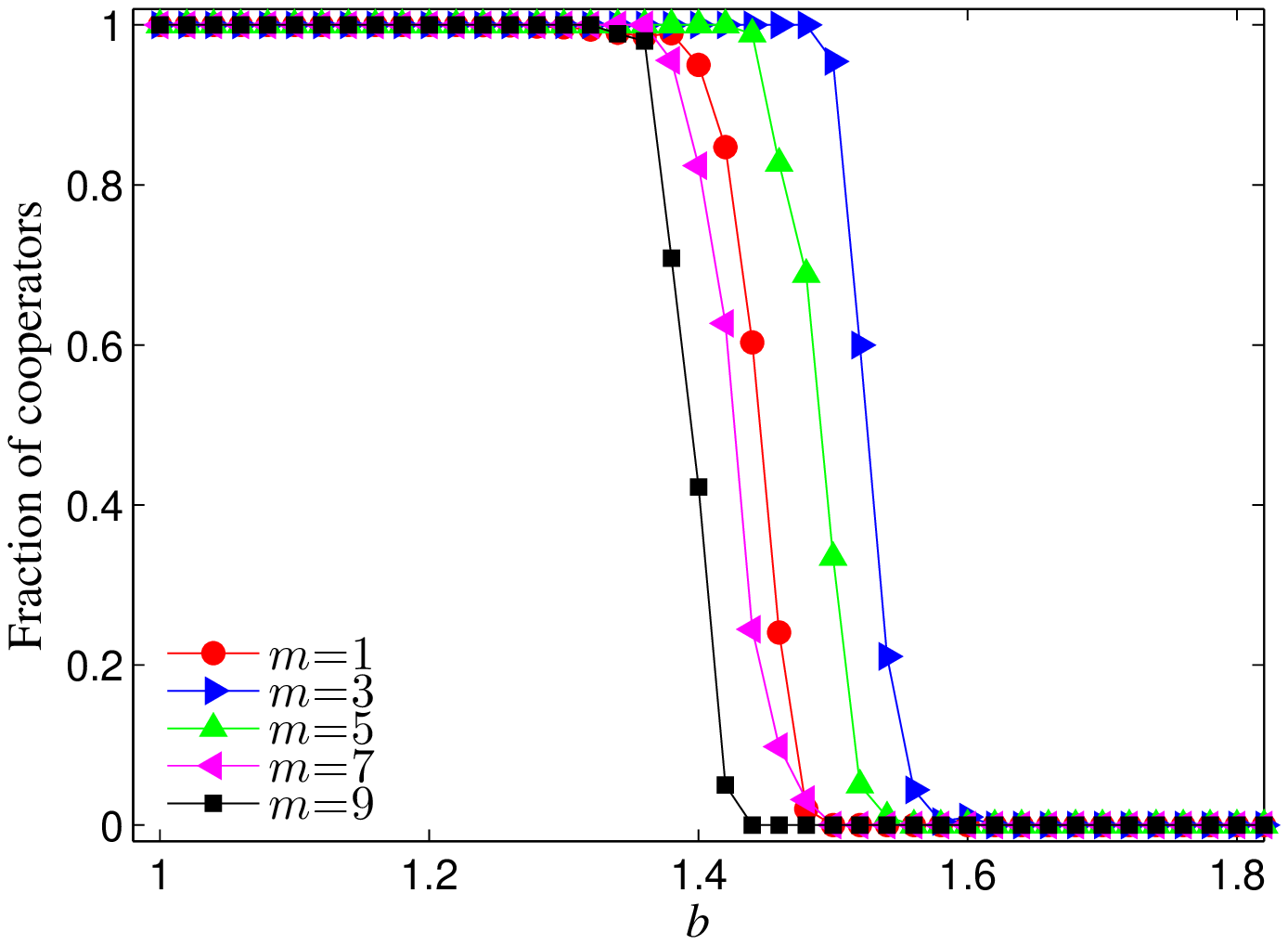}\end{center}
\caption{\textbf{Fraction of cooperators in dependence on the temptation to defect $b$ for different values of $m$.} It can be observed that intermediate values of $m$ are optimal for the evolution of cooperation, albeit this depends somewhat on the temptation to defect $b$. Presented results are averages over 100 independent realizations obtained with the system size $N=10^4$ and $\lambda=0.99$. Lines connecting the symbols are just to guide the eye.}
\label{fig1}
\end{figure}

\begin{figure}
\begin{center}\includegraphics[width=9cm]{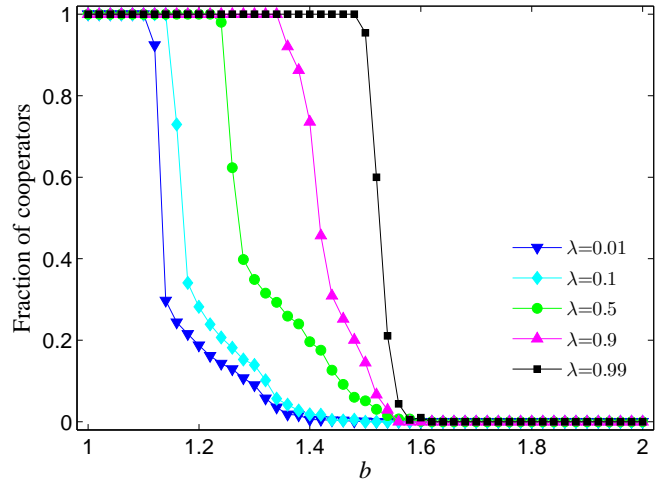}\end{center}
\caption{\textbf{Fraction of cooperators in dependence on the temptation to defect $b$ for different values of $\lambda$.} It can be observed that the higher the $\lambda$ the larger the temptation to defect $b$ at which cooperators are able to survive when competing against defectors. The span of $b$ values where cooperators are able to dominate completely increases as well with increasing $\lambda$. Presented results are averages over 100 independent realizations obtained with the system size $N=10^4$ and $m=2$. Lines connecting the symbols are just to guide the eye.}
\label{fig2}
\end{figure}

\begin{figure}
\begin{center}\includegraphics[width=9cm]{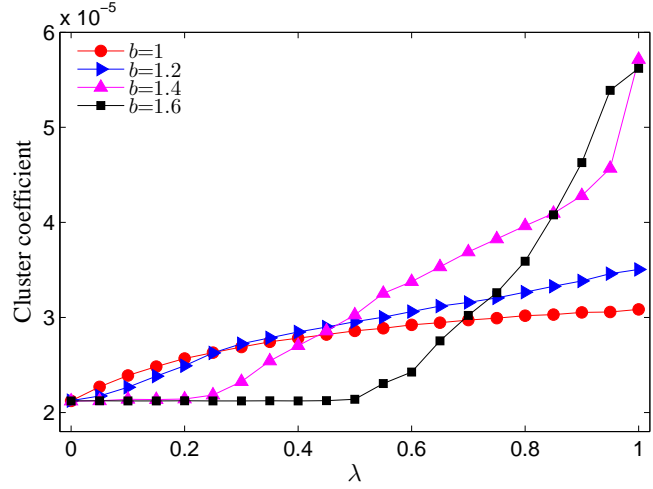}\end{center}
\caption{\textbf{Clustering coefficient of the resulting networks in dependence on $\lambda$ for different values of the temptation to defect $b$.} It can be observed that larger values of $\lambda$ in general lead to more clustered networks, and that higher $b$ promote clustering as well. Presented results are averages over 100 independent realizations obtained with the system size $N=10^4$ and $m=3$. Lines connecting the symbols are just to guide the eye.}
\label{fig3}
\end{figure}

\begin{figure}
\begin{center}\includegraphics[width=9cm]{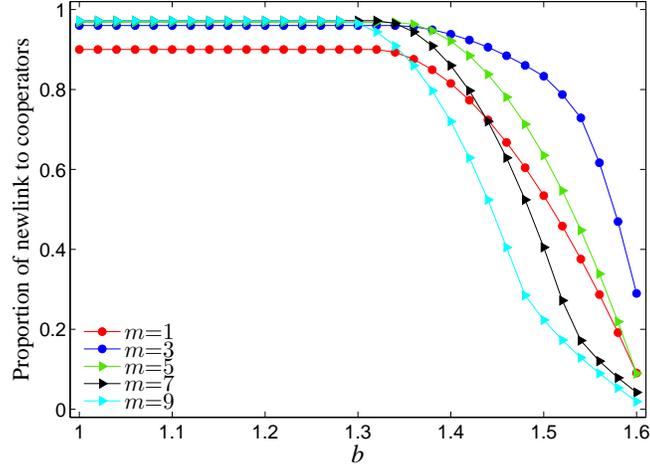}\end{center}
\caption{\textbf{Fraction of new links that are assigned to cooperators in dependence on the temptation to defect $b$ for different values of $m$.} It can be observed that the higher the temptation to defect $b$, the lower the fraction of new links that are received by cooperators. As by results presented in Fig.~\ref{fig1}, it can be concluded that intermediate values of $m$ are optimal for cooperators to expand their neighborhoods, although as before, here too this depends somewhat on the temptation to defect $b$. Altogether, this leads to the conclusion that who (either cooperators or defectors) obtains the new links is crucial for the successful evolution of cooperation. Presented results are averages over 100 independent realizations obtained with the system size $N=10^4$ and $\lambda=0.99$. Lines connecting the symbols are just to guide the eye.}
\label{fig4}
\end{figure}

\begin{figure}
\begin{center}\includegraphics[width=9cm]{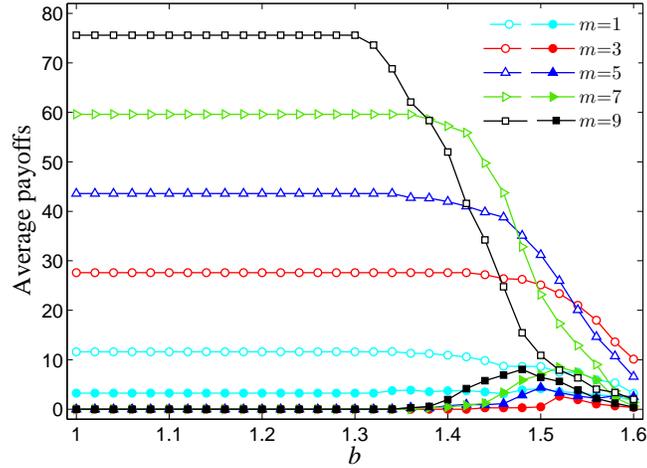}\end{center}
\caption{\textbf{Average payoffs of cooperators (open symbols) and defectors (filled symbols) in dependence on the temptation to defect $b$ for different values of $m$.} The success of different values of $m$ to optimally promote the evolution of cooperation is reflected also in the average payoffs, with intermediate values of $m$ clearly maintaining cooperators more successful than defectors even at high values of $b$. To a lesser extent this is true for small (\textit{e.g.} $m=1$) and large (\textit{e.g.} $m=9$) values of $m$, although for small values of $b$ higher values of $m$ are actually the most effective. The optimal value of $m$ thus depends on the severity of the social dilemma. While low temptations to defect are offset more effectively by larger values of $m$, high temptations to defect are dealt with better by intermediate values of $m$ (note that at $b=1.6$ the intermediate value $m=3$ warrants the biggest difference between the average payoffs of the two strategies). Presented results are averages over 100 independent realizations obtained with the system size $N=10^4$ and $\lambda=0.99$. Lines connecting the symbols are just to guide the eye.}
\label{fig5}
\end{figure}

\begin{figure}
\begin{center}\includegraphics[width=9cm]{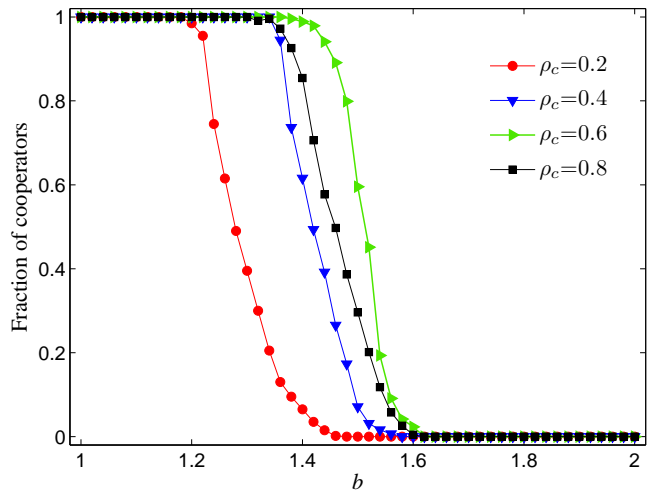}\end{center}
\caption{\textbf{Fraction of cooperators in dependence on the temptation to defect $b$ for different initial fractions of cooperators $\rho_{c}$.} It is interesting to observe that too high initial values of $\rho_{c}$ may act detrimental on the evolution of cooperation in the considered model. This may be attributed to the fact that defectors thrive in populations where there are numerous cooperators to exploit, and ultimately this may become a disadvantage in the latter stages of the game. Presented results indicate that an intermediate initial level of cooperators is optimal for the evolution of cooperation. Presented results are averages over 100 independent realizations obtained with the system size $N=10^4$, $m=3$ and $\lambda=0.99$. Lines connecting the symbols are just to guide the eye.}
\label{fig6}
\end{figure}


\begin{thebibliography}{10}
\providecommand{\url}[1]{\texttt{#1}}
\providecommand{\urlprefix}{URL }
\expandafter\ifx\csname urlstyle\endcsname\relax
  \providecommand{\doi}[1]{doi:\discretionary{}{}{}#1}\else
  \providecommand{\doi}{doi:\discretionary{}{}{}\begingroup
  \urlstyle{rm}\Url}\fi
\providecommand{\bibAnnoteFile}[1]{%
  \IfFileExists{#1}{\begin{quotation}\noindent\textsc{Key:} #1\\
  \textsc{Annotation:}\ \input{#1}\end{quotation}}{}}
\providecommand{\bibAnnote}[2]{%
  \begin{quotation}\noindent\textsc{Key:} #1\\
  \textsc{Annotation:}\ #2\end{quotation}}
\providecommand{\eprint}[2][]{\url{#2}}

\bibitem{axelrod_84}
Axelrod R (1984) The Evolution of Cooperation.
\newblock New York: Basic Books.
\input{axelrod_84}

\bibitem{sigmund_93}
Sigmund K (1993) Games of Life: Exploration in Ecology, Evolution and Behavior.
\newblock Oxford, UK: Oxford University Press.
\input{sigmund_93}

\bibitem{weibull_95}
Weibull JW (1995) Evolutionary Game Theory.
\newblock Cambridge, MA: MIT Press.
\input{weibull_95}

\bibitem{hofbauer_98}
Hofbauer J, Sigmund K (1998) Evolutionary Games and Population Dynamics.
\newblock Cambridge, UK: Cambridge University Press.
\input{hofbauer_98}

\bibitem{gintis_00}
Gintis H (2000) Game Theory Evolving.
\newblock Princeton: Princeton University Press.
\input{gintis_00}

\bibitem{nowak_06}
Nowak MA (2006) Evolutionary Dynamics.
\newblock Cambridge, MA: Harvard University Press.
\input{nowak_06}

\bibitem{sigmund_10}
Sigmund K (2010) The Calculus of Selfishness.
\newblock Princeton, MA: Princeton University Press.
\input{sigmund_10}

\bibitem{doebeli_el05}
Doebeli M, Hauert C (2005) Models of cooperation based on prisoner's dilemma
  and snowdrift game.
\newblock Ecol Lett 8: 748--766.
\input{doebeli_el05}

\bibitem{nowak_s06}
Nowak MA (2006) Five rules for the evolution of cooperation.
\newblock Science 314: 1560--1563.
\input{nowak_s06}

\bibitem{szabo_pr07}
Szab{\'o} G, F{\'a}th G (2007) Evolutionary games on graphs.
\newblock Phys Rep 446: 97--216.
\input{szabo_pr07}

\bibitem{schuster_jbp08}
Schuster S, Kreft JU, Schroeter A, Pfeiffer T (2008) Use of game-theoretical
  methods in biochemistry and biophysics.
\newblock J Biol Phys 34: 1--17.
\input{schuster_jbp08}

\bibitem{perc_bs10}
Perc M, Szolnoki A (2010) Coevolutionary games -- a mini review.
\newblock BioSystems 99: 109--125.
\input{perc_bs10}

\bibitem{nowak_n92b}
Nowak MA, May RM (1992) Evolutionary games and spatial chaos.
\newblock Nature 359: 826--829.
\input{nowak_n92b}

\bibitem{hauert_n04}
Hauert C, Doebeli M (2004) Spatial structure often inhibits the evolution of
  cooperation in the snowdrift game.
\newblock Nature 428: 643--646.
\input{hauert_n04}

\bibitem{hamilton_wd_jtb64a}
Hamilton WD (1964) Genetical evolution of social behavior \protect{I}.
\newblock J Theor Biol 7: 1--16.
\input{hamilton_wd_jtb64a}

\bibitem{axelrod_s81}
Axelrod R, Hamilton WD (1981) The evolution of cooperation.
\newblock Science 211: 1390--1396.
\input{axelrod_s81}

\bibitem{ohtsuki_jtb04}
Ohtsuki H (2004) Reactive strategies in indirect reciprocity.
\newblock J Theor Biol 227: 299--314.
\input{ohtsuki_jtb04}

\bibitem{ohtsuki_jtb06b}
Ohtsuki H, Iwasa Y (2006) The leading eight: Social norms that can maintain
  cooperation by indirect reciprocity.
\newblock J Theor Biol 239: 435--444.
\input{ohtsuki_jtb06b}

\bibitem{brandt_jtb06}
Brandt H, Sigmund K (2006) The good, the bad and the disciminator - errors in
  direct and indirect reciprocity.
\newblock J Theor Biol 239: 183--194.
\input{brandt_jtb06}

\bibitem{pacheco_jtb08}
Pacheco JM, Traulsen A, Ohtsuki H, Nowak MA (2008) Repeated games and direct
  reciprocity under active linking.
\newblock J Theor Biol 250: 723--731.
\input{pacheco_jtb08}

\bibitem{wilson_ds_an77}
Wilson DS (1977) Structured demes and the evolution of group-advantageous
  traits.
\newblock Am Nat 111: 157--185.
\input{wilson_ds_an77}

\bibitem{dugatkin_bs96}
Dugatkin LA, Mesterton-Gibbons M (1996) Cooperation among unrelated
  individuals: reciprocial altruism, by-product mutualism and group selection
  in fishes.
\newblock Biosystems 37: 19--30.
\input{dugatkin_bs96}

\bibitem{traulsen_pnas06}
Traulsen A, Nowak MA (2006) Evolution of cooperation by multilevel selection.
\newblock Proc Natl Acad Sci USA 103: 10952--10955.
\input{traulsen_pnas06}

\bibitem{szolnoki_njp09}
Szolnoki A, Perc M (2009) Emergence of multilevel selection in the prisoner's
  dilemma game on coevolving random networks.
\newblock New J Phys 11: 093033.
\input{szolnoki_njp09}

\bibitem{abramson_pre01}
Abramson G, Kuperman M (2001) Social games in a social network.
\newblock Phys Rev E 63: 030901(R).
\input{abramson_pre01}

\bibitem{ebel_pre02}
Ebel H, Bornholdt S (2002) Coevolutionary games on networks.
\newblock Phys Rev E 66: 056118.
\input{ebel_pre02}

\bibitem{holme_pre03}
Holme P, Trusina A, Kim BJ, Minnhagen P (2003) Prisoner's dilemma in real-world
  acquaintance networks: Spikes and quasiequilibria induced by the interplay
  between structure and dynamics.
\newblock Phys Rev E 68: 030901.
\input{holme_pre03}

\bibitem{wu_zx_pre05}
Wu ZX, Xu XJ, Chen Y, Wang YH (2005) Spatial prisoner's dilemma game with
  volunteering in \protect{Newman-Watts} small-world networks.
\newblock Phys Rev E 71: 037103.
\input{wu_zx_pre05}

\bibitem{tomassini_pre06}
Tomassini M, Luthi L, Giacobini M (2006) Hawks and doves games on small-world
  networks.
\newblock Phys Rev E 73: 016132.
\input{tomassini_pre06}

\bibitem{ohtsuki_n06}
Ohtsuki H, Hauert C, Lieberman E, Nowak MA (2006) A simple rule for the
  evolution of cooperation on graphs and social networks.
\newblock Nature 441: 502--505.
\input{ohtsuki_n06}

\bibitem{ren_pre07}
Ren J, Wang WX, Qi F (2007) Randomness enhances cooperation: coherence
  resonance in evolutionary game.
\newblock Phys Rev E 75: 045101(R).
\input{ren_pre07}

\bibitem{wu_zx_pa07}
Wu ZX, Guan JY, Xu XJ, Y-HWang (2007) Evolutionary prisoner's dilemma game on
  \protect{Barab{\'a}si-Albert} scale-free networks.
\newblock Physica A 379: 672--680.
\input{wu_zx_pa07}

\bibitem{vukov_pre08}
Vukov J, Szab{\'o} G, Szolnoki A (2008) Prisoner's dilemma game on
  \protect{Newman-Watts} graphs.
\newblock Phys Rev E 77: 026109.
\input{vukov_pre08}

\bibitem{luthi_pa08}
Luthi L, Pestelacci E, Tomassini M (2008) Cooperation and community structure
  in social networks.
\newblock Physica A : 955--966.
\input{luthi_pa08}

\bibitem{floria_pre09}
Flor{\'{\i}}a LM, Gracia-L{\'a}zaro C, G{\'o}mez-Garde{\~n}es J, Moreno Y
  (2009) Social network reciprocity as a phase transition in evolutionary
  cooperation.
\newblock Phys Rev E 79: 026106.
\input{floria_pre09}

\bibitem{poncela_epl09}
Poncela J, G{\'o}mez-Garde{\~n}es J, Flor{\' \i}a LM, Moreno Y, S{\'a}nchez A
  (2009) Cooperative scale-free networks despite the presence of defector hubs.
\newblock EPL 88: 38003.
\input{poncela_epl09}

\bibitem{santos_prl05}
Santos FC, Pacheco JM (2005) Scale-free networks provide a unifying framework
  for the emergence of cooperation.
\newblock Phys Rev Lett 95: 098104.
\input{santos_prl05}

\bibitem{santos_pnas06}
Santos FC, Pacheco JM, Lenaerts T (2006) Evolutionary dynamics of social
  dilemmas in structured heterogeneous populations.
\newblock Proc Natl Acad Sci USA 103: 3490--3494.
\input{santos_pnas06}

\bibitem{santos_jeb06}
Santos FC, Pacheco JM (2006) A new route to the evolution of cooperation.
\newblock J Evol Biol 19: 726--733.
\input{santos_jeb06}

\bibitem{gomez-gardenes_prl07}
G{\'o}mez-Garde{\~n}es J, Campillo M, Moreno Y, Flor{\' \i}a LM (2007)
  Dynamical organization of cooperation in complex networks.
\newblock Phys Rev Lett 98: 108103.
\input{gomez-gardenes_prl07}

\bibitem{poncela_njp07}
Poncela J, G{\'o}mez-Garde{\~n}es J, Flor{\' \i}a LM, Moreno Y (2007)
  Robustness of cooperation in the evolutionary prisoner's dilemma on complex
  systems.
\newblock New J Phys 9: 184.
\input{poncela_njp07}

\bibitem{szolnoki_pa08}
Szolnoki A, Perc M, Danku Z (2008) Towards effective payoffs in the prisoner's
  dilemma game on scale-free networks.
\newblock Physica A 387: 2075--2082.
\input{szolnoki_pa08}

\bibitem{gomez-gardenes_jtb08}
G{\'o}mez-Garde{\~n}es J, Poncela J, Flor{\' \i}a LM, Moreno Y (2008) Natural
  selection of cooperation and degree hierarchy in heterogeneous populations.
\newblock J Theor Biol 253: 296--301.
\input{gomez-gardenes_jtb08}

\bibitem{wu_zx_pre06}
Wu ZX, Xu XJ, Huang ZG, Wang SJ, Wang YH (2006) Evolutionary prisoner's dilemma
  game with dynamic preferential selection.
\newblock Phys Rev E 74: 021107.
\input{wu_zx_pre06}

\bibitem{kim_bj_pre02}
Kim BJ, Trusina A, Holme P, Minnhagen P, Chung JS, et~al. (2002) Dynamic
  instabilities induced by asymmetric influence: Prisoner's dilemma game in
  small-world networks.
\newblock Phys Rev E 66: 021907.
\input{kim_bj_pre02}

\bibitem{szolnoki_epl07}
Szolnoki A, Szab{\'o} G (2007) Cooperation enhanced by inhomogeneous activity
  of teaching for evolutionary prisoner's dilemma games.
\newblock EPL 77: 30004.
\input{szolnoki_epl07}

\bibitem{wu_zx_cpl06}
Wu ZX, Xu XJ, Wang YH (2006) Prisoner's dilemma game with heterogeneous
  influental effect on regular small-world networks.
\newblock Chin Phys Lett 23: 531--534.
\input{wu_zx_cpl06}

\bibitem{perc_pre08}
Perc M, Szolnoki A (2008) Social diversity and promotion of cooperation in the
  spatial prisoner's dilemma game.
\newblock Phys Rev E 77: 011904.
\input{perc_pre08}

\bibitem{wu_zx_pre09b}
Wu ZX, Rong Z, Holme P (2009) Diversity of reproduction time scale promotes
  cooperation in spatial prisoner's dilemma games.
\newblock Phys Rev E 80: 036103.
\input{wu_zx_pre09b}

\bibitem{albert_rmp02}
Albert R, Barab{\'a}si AL (2002) Statistical mechanics of complex networks.
\newblock Rev Mod Phys 74: 47--97.
\input{albert_rmp02}

\bibitem{newman_siamr03}
Newman MEJ (2003) The structure and function of complex networks.
\newblock SIAM Review 45: 167--256.
\input{newman_siamr03}

\bibitem{dorogovtsev_03}
Dorogovtsev SN, Mendes JFF (2003) Evolution of Networks: From Biological Nets
  to the Internet and \protect{WWW}.
\newblock Oxford: Oxford Univ. Press.
\input{dorogovtsev_03}

\bibitem{boccaletti_pr06}
Boccaletti S, Latora V, Moreno Y, Chavez M, Hwang D (2006) Complex networks:
  Structure and dynamics.
\newblock Phys Rep 424: 175--308.
\input{boccaletti_pr06}

\bibitem{pacheco_prl06}
Pacheco JM, Traulsen A, Nowak MA (2006) Coevolution of strategy and structure
  in complex networks with dynamical linking.
\newblock Phys Rev Lett 97: 258103.
\input{pacheco_prl06}

\bibitem{zimmermann_pre04}
Zimmermann MG, Egu{\'{\i}}luz V, Miguel MS (2004) Coevolution of dynamical
  states and interactions in dynamic networks.
\newblock Phys Rev E 69: 065102(R).
\input{zimmermann_pre04}

\bibitem{zimmermann_pre05}
Zimmermann MG, Egu{\'{\i}}luz V (2005) Cooperation, social networks and the
  emergence of leadership in a prisoner's dilemma with local interactions.
\newblock Phys Rev E 72: 056118.
\input{zimmermann_pre05}

\bibitem{perc_njp06c}
Perc M (2006) Double resonance in cooperation induced by noise and network
  variation for an evolutionary prisoner's dilemma.
\newblock New J Phys 8: 183.
\input{perc_njp06c}

\bibitem{szolnoki_epl09}
Szolnoki A, Perc M (2009) Resolving social dilemmas on evolving random
  networks.
\newblock EPL 86: 30007.
\input{szolnoki_epl09}

\bibitem{fu_pre09}
Fu F, Wu T, Wang L (2009) Partner switching stabilizes cooperation in
  coevolutionary prisoner's dilemma.
\newblock Phys Rev E 79: 036101.
\input{fu_pre09}

\bibitem{wu_b_pone10}
Wu B, Zhou D, Fu F, Luo Q, Wang L, et~al. (2010) Evolution of cooperation on
  stochastic dynamical networks.
\newblock PLoS ONE 5: e11187.
\input{wu_b_pone10}

\bibitem{van-segbroeck_bmceb08}
Van~Segbroeck S, Santos FC, Now{\'e} A, Pacheco JM, Lenaerts T (2008) The
  evolution of prompt reaction to adverse ties.
\newblock BMC Evolutionary Biolology 8: 287--294.
\input{van-segbroeck_bmceb08}

\bibitem{van-segbroeck_prl09}
Van~Segbroeck S, Santos FC, Lenaerts T, Pacheco JM (2009) Reacting differently
  to adverse ties promotes cooperation in social networks.
\newblock Phys Rev Lett 102: 058105.
\input{van-segbroeck_prl09}

\bibitem{poncela_ploso08}
Poncela J, G{\'o}mez-Garde{\~n}es J, Flor{\' \i}a LM, S\'anchez A, Moreno Y
  (2008) Complex cooperative networks from evolutionary preferential
  attachment.
\newblock PLoS ONE 3: e2449.
\input{poncela_ploso08}

\bibitem{poncela_njp09}
Poncela J, G{\'o}mez-Garde{\~n}es J, Traulsen A, Moreno Y (2009) Evolutionary
  game dynamics in a growing structured population.
\newblock New J Phys 11: 083031.
\input{poncela_njp09}

\bibitem{lee_s_prl11}
Lee S, Holme P, Wu ZX (2011) Emergent hierarchical structures in multiadaptive
  games.
\newblock Phys Rev Lett 106: 028702.
\input{lee_s_prl11}

\bibitem{vainstein_jtb07}
Vainstein MH, Silva ATC, Arenzon JJ (2007) Does mobility decrease cooperation?
\newblock J Theor Biol 244: 722--728.
\input{vainstein_jtb07}

\bibitem{szolnoki_njp08}
Szolnoki A, Perc M (2008) Coevolution of teaching activity promotes
  cooperation.
\newblock New J Phys 10: 043036.
\input{szolnoki_njp08}

\bibitem{szolnoki_epjb09}
Szolnoki A, Perc M (2009) Promoting cooperation in social dilemmas via simple
  coevolutionary rules.
\newblock Eur Phys J B 67: 337--344.
\input{szolnoki_epjb09}

\bibitem{droz_epjb09}
Droz M, Szwabinski J, Szab{\'o} G (2009) Motion of influential players can
  support cooperation in prisoner's dilemma.
\newblock Eur Phys J B 71: 579--585.
\input{droz_epjb09}

\bibitem{wardil_epl09}
Wardil L, da~Silva JKL (2009) Adoption of simultaneous different strategies
  against different opponents enhances cooperation.
\newblock EPL 86: 38001.
\input{wardil_epl09}

\bibitem{wardil_pre10}
Wardil L, sa~Silva JKL (2010) Distinguishing the opponents promotes cooperation
  in well-mixed populations.
\newblock Phys Rev E 81: 036115.
\input{wardil_pre10}

\bibitem{szabo_epl09}
Szab{\'o} G, Szolnoki A, Vukov J (2009) Selection of dynamical rules in spatial
  prisoner's dilemma games.
\newblock EPL 87: 18007.
\input{szabo_epl09}

\bibitem{szolnoki_pre09}
Szolnoki A, Perc M, Szab{\'o} G, Stark HU (2009) Impact of aging on the
  evolution of cooperation in the spatial prisoner's dilemma game.
\newblock Phys Rev E 80: 021901.
\input{szolnoki_pre09}

\bibitem{szolnoki_pre09d}
Szolnoki A, Vukov J, Szab{\'o} G (2009) Selection of noise level in strategy
  adoption for spatial social dilemmas.
\newblock Phys Rev E 80: 056112.
\input{szolnoki_pre09d}

\bibitem{sicardi_jtb09}
Sicardi EA, Fort H, Vainstein MH, Arenzon JJ (2009) Random mobility and spatial
  structure often enhance cooperation.
\newblock J Theor Biol 256: 240--246.
\input{sicardi_jtb09}

\bibitem{tanimoto_pa09}
Tanimoto J (2009) Promotion of cooperation through co-evolution of networks and
  strategy in a $2 \times 2$ game.
\newblock Physica A 388: 953--960.
\input{tanimoto_pa09}

\bibitem{cao_l_jtb11}
Cao L, Ohtsuki H, Wang B, Aihara K (2011) Evolution of cooperation on
  adaptively weighted networks.
\newblock J Theor Biol 272: 8--15.
\input{cao_l_jtb11}

\bibitem{szolnoki_epl08}
Szolnoki A, Perc M, Danku Z (2008) Making new connections towards cooperation
  in the prisoner's dilemma game.
\newblock EPL 84: 50007.
\input{szolnoki_epl08}

\bibitem{szabo_pre98}
Szab{\'o} G, T{\H{o}}ke C (1998) Evolutionary prisoner's dilemma game on a
  square lattice.
\newblock Phys Rev E 58: 69--73.
\input{szabo_pre98}

\bibitem{roca_prl06}
Roca CP, Cuesta JA, S{\'a}nchez A (2006) Time scales in evolutionary dynamics.
\newblock Phys Rev Lett 97: 158701.
\input{roca_prl06}

\end{thebibliography}
\end{document}